# How the University Portal Inspired Changes in the Academic Assessment Culture

Valerii Semenets [1], Svitlana Gryshko [2], Mariia Golovianko [3], Oleksandr Shevchenko [3], Liudmyla Titova [3], Olena Kaikova [4], Vagan Terziyan [4*], Timo Tiihonen [4]

[1] Department of Metrology and Technical Expertise, Kharkiv National University of Radioelectronics, Ukraine
[2] Department of Economic Cybernetics and Management of Economic Security, Kharkiv National University of Radioelectronics, Ukraine
[3] Department of Artificial Intelligence, Kharkiv National University of Radioelectronics, Ukraine
[4] Faculty of Information Technology, University of Jyvaskyla, Finland

## ABSTRACT

Information retrieval (IR) is known facilitator of changes ongoing in human society and vice versa. This is due to the fact that IR is a key component of the digital ecosystems, where both information providers and information consumers collaboratively address their problems with the use of technologies. Organization and design of such ecosystems drives particular social impact for all the players involved. In this paper, we study the impact made by a particular IR ecosystem (semantic portal) used for management of academic information resources and processes within the Ukrainian higher education. We show how this portal is changing a collective mindset of the academic community of its users. We argue that such impact becomes possible due to specific organization of the ecosystem, where all the information resources, IR services and related analytics (search, assessment, ranking, etc.) and IR users inhabit the same semantic space under umbrella of the corresponding ontologies. Personal values and preferences of the users configure on-the-fly the corresponding IR analytics and enable personalized value-driven IR services, making everyone feel involved into the organizational decision-making processes. Four years of active use of this portal in university environment has been reported and related impact is evaluated in this study.

## 1 Introduction

Information Retrieval (IR) tools, systems and services, like many other forms of technology, are not socially neutral. IR affects and is itself affected by society. The evolution of the way we search and use information is changing also the way people coevolve within their business and living environments. The collective minds of the society, in turn, is influencing the development of IR through the needs people have for processing information. Various social aspects of the collaborative IR has been pointed out quite a while ago (Karamuftuoglu, 1998), where it is argued that the fundamental for IR production and consumption of knowledge is deeply embedded in the practices of a users' community. The emergence of social informatics for learning about social aspects of information-seeking and applying that knowledge to the design, development and evaluation of systems has been discussed in (Tang, 2000). There is also a concern related to IR within information resources with polarized opinions about same issues as the political bias in the top search results can play a significant role in shaping public opinion towards certain perspectives (Kulshrestha, 2019). Since the patent (Dasan, 1998) related to the personalized IR based on user-defined profile, an important aspect has been added to the IR domain, i.e., including the user to the IR loop. Organizing information (when possible) with the help of semantic graphs and ontologies essentially improve the performance of the IR towards information extraction and inference (Mayfield & Finin, 2003). Exploratory search (Marchionini, 2006) turns the IR to be the instrument for serious cognitive creative activities of the users including discovery of hidden opportunities during the search practices. Other modern scenarios of interacting (online vs. offline) with the IR has been discussed in (Markov & Rijke, 2019), and many unexplored opportunities to improve the interaction process has been pointed out.

In this paper we provide a concrete example of a sophisticated IR solution (with some "social flavor") and the surprising impact it made to the community of its users due to unique design architecture. The rest of the paper is organized as follows: In Section 2, we describe the object of study, i.e. the value-driven IR ecosystem for management of the Ukrainian higher education; in Section 3, we describe how this ecosystem changes the collective mindset of the Ukrainian academic community; in Section 4, we briefly review related work; and we conclude in Section 5.

## 2 Semantic portal as a value-driven information retrieval ecosystem

The object of our study is a semantic Web-based portal (Terziyan et al., 2015) designed for storage, retrieval and assessment of various "valuable" information resources. Resources are organized as a semantic graph under umbrella of the domain ontology. Stored resources are the subject of retrieval and assessment based on their properties and driven by embedded analytics. The analytics is organized so that it can support various options for chosen domain and, therefore, all the analytical functions has been represented as semantic services under umbrella of the domain-independent service ontology.

---

[*] Contact author: vagan.terziyan@jyu.fi

The ecosystem around the portal assumes that the contributors (resource providers), who are also the users (resource retrievers), register at the portal as special types of semantic resources. This process is driven by special ontology (ontology of personal values). Important here is that (when registering) everyone positions oneself in the semantic space of the value system (values for personal preferences regarding importance of each particular feature for future evaluation of the different information resources' categories). Such personal value systems work by autonomously representing opinions of everyone when the evaluation of a particular resource (or ranking of the resource group) is being made. Each value system in fact contains values for the variables used by the portal embedded analytical services. Similar registration (with the value system provision) is possible also for organizational entities and is supported by the ontology of organizational values.

Such organization of the ecosystem around the portal enables several important IR service options:

- smart executable IR queries, when the query automatically configures and executes also the analytical functions (regarding the value system of a particular user) and, as a result, the user gets not only the needed resource itself but also various assessments and analytical reports around it adapted to the user preferences and even to the know user emotional state. Details on the underlying approaches ("executable reality" and "emotional business intelligence") and the underlying technology ("knowledge computing") are reported in (Terziyan & Kaykova, 2012; Terziyan & Kaykova, 2015) and in (Terziyan et al., 2014);
- ranking IR queries, when the group of resources is discovered on the basis of the query, each of the resources is automatically assessed driven by user references and the scored ranking list of the resources is delivered to the user (Terziyan et al., 2015);
- pragmatic and exploratory IR queries (smart decision-support), when the query is first processed to discover the hidden user intentions (e.g., what kind of a decision the user is probably going to make on the basis of retrieved content), then the analytics (personalized according to the user value system) assesses each of possible decision options and, finally, the ranking list of the decision options is provided to the user (Terziyan et al., 2017);
- IR queries driven by collective search experiences, when the search-history-aware association-driven navigation across information spaces is supported in spite of different value systems of the users who assessed the information resources in question before (Golovianko, 2018).

As a pilot for design, implementation and testing of such semantic-portal-driven IR ecosystem, we have chosen the higher education domain and appropriate valuable information resources (academic achievements) and the users (academic staff and public). The rest of the paper is dedicated to the evaluation and reporting the visible impact of the IR ecosystem as a collective awareness platform (TRUST Portal: https://portal.dovira.eu/) for higher education after four years of active use.

## 3 Introducing the collective awareness platform to transform the collective academic mindset

Universities are powerful drivers of societal development as they provide knowledge for positive transformations. Hence, higher education, in order to keep up with the digital disruption, globalization, and other contemporary challenges, as well as to be truly transformative, must constantly transform also itself (COPERNICUS, 2015). For highly ranked universities in the developed countries (Ehlers & Kellermann, 2019), design of effective learning environments is a high priority task, as it should be also for developing countries, such as Ukraine, that struggle with the inheritance of the old closed highly centralized soviet system. Having started education reforms and joining the Bologna Process in 2005, Ukraine expected quick improvements in research capabilities, international presence, quality of teaching and student experiences. Very soon, it became obvious: even the most effective and reformist solutions will not work unless the academic community is ready to accept them. Ukrainian academic community appeared to be unprepared to the new realities and resisting strongly any changes. Long history of isolation, low level of financing, absence of academic freedom and institutional autonomy played a cruel joke with Ukrainian HEIs – intended reforms led to catastrophic depreciation of valuable academic results and booming corruption culture. Ukrainian HE urgently needed new instruments to trigger a change towards new academic values and quality culture (Golovianko et al., 2013).

TRUST Portal of HE Quality Assurance (http://portal.dovira.eu) was initially developed in a TEMPUS project[1] as an open web environment for IR of academic resources, digital twinning of HE processes to increase transparency and trust (Terziyan et al., 2015). With the Portal, academics could publicly report and store their academic achievements, search heterogeneous academic information, verify the truthfulness of the reports and check the progress of the competitors. Moreover, now everyone interested in the assessments and analytics of HE could calculate the impact of academic results using different measures or rank HE players, institutions, units, and organizations according to own preferred quality indicators on the fly. These customization capabilities and the embedded flexibility were enabled by the semantic state-of-the-art technologies, such as the ontology-based organization of the data storage and knowledge inference mechanisms on top of it.

Transparency was not the only beneficial feature supported by the Portal: being designed as a collective intelligence platform according to the user-centered social media principles, the Portal could provoke a bottom-up public-initiated and public-driven activity in Ukrainian HEIs. For the first time academics could experience the enrolment into the actual decision-making. This triggered a strong motivation for change and development. Sharing the best practices and personal/organizational values systems (quality indicators) facilitated benchmarking, collective wisdom and elaboration of quality culture.

---

[1] Project TRUST (#516935-TEMPUS-1-2011-1-FITEMPUS-SMGR)

The Portal was introduced on the national level during the election of the representatives of private HE to the first independent Ukrainian Quality Assurance Agency in 2015. The candidates were subjected to transparent Portal-supported ranking according to the formal requirements declared by the by the public officials. The ranking results contradicted the results of the formal political nomination of members, which then caused a negative public response and along with other factors forced the Ministry to revise both the unfair results and the selection process.

The first long-term pilot of the Portal was run in Kharkiv National University of Radioelectronics (NURE) (where it was developed). After the first launch in 2014 it was used mainly by a small group of open-minded academics involved into international collaboration. A large part of the university community, especially those with poor academic results, was not motivated to contribute information to the Portal, considering it as an additional unnecessary workload. Moreover, internal disagreements in NURE along with the political volatility inhibited the full-scale introduction of the Portal for several years. Quite ironically, the Portal was finally adopted by managers and academics only after a critical mass of change agents emerged in the administration of the university. The newly elected rector supported the movement from corruption towards elaboration of quality culture by introducing the Portal into university decision making. In the beginning of 2018, in order to promote the new academic values and the Portal among the university staff, NURE Academic Council announced a contest for the researchers: considerable bonuses were proposed to those on top of the Portal ranking lists created according to the internationally recognized criteria, such as citation indices, publications in journals with high impact factor, impact creating participation in international collaborations, etc. The criteria were published beforehand in April and both young and experienced researchers were awarded publicly in December.

## 4 Related work

Our case study provides practical justification for human development and modernization theories, illustrating the role of IR in transforming the collective mindset. Democratization advances human development, it enlarges people's freedoms and opportunities. However, formal empowerment is not a guarantee for the proper use of the obtained rights (Boucher, 2009). Democracy contributes to actual human development when supported by cultural traditions that emphasize human social values that favor liberalization and public motivated engagement into decision-making (Inglehart & Welzel, 2010). Public enrolment into changes is very important for Organization Development (Cummings & Worley, 2014), especially in higher education (European University Association, 2006). Thus, a "bottom-up" pro-democratic civic culture should be elaborated to augment the existing "top-down" managerial approaches (Osseo-Asare & Pieris, 2007). In higher education, it means a shift from quality control to increased autonomy, credibility and educational enhancement based on the experiences, expertise and values. It implies collective responsibility and engagement for all community members, including students, academic and administrative staff based on mutual trust, the importance of which has been widely recognized in this matter (Leveille, 2006; Dzimińska et al., 2018).

IR-driven e-democracy tools, such as open collective awareness platforms, can help to foster more active and responsible bottom-up activity by promoting, ensuring and enhancing needed transparency, accountability, responsiveness, engagement, deliberation, inclusiveness, accessibility, participation and social cohesion. This phenomenon can be attributed to the paradigm of Cultural Computing (Rauterberg, 2006), and is used widely, particularly in a new field of study of Collective Intelligence enabled by Cognitive Computing[2]. It is caused by the following aspects of social interaction supported by ICT tools: transparent and unbiased decision-making increases trust (Newton et al., 2018); processes run at collective platforms contribute to building of the social relationships, thus, social capital (Lin, 2017); public "rules of the game", i.e., values system, public recognition of achievements, transparency and accessibility of procedures create conditions for the distributive, procedural and interactional justice (Virtanen & Elovanio, 2018).

## 5 Conclusions

Collaborative IR platforms, such as TRUST Portal, do not only ensure educational processes "as usual", they serve as open environments for the collection of the needed experimental evidence in favor of the developmental theories promoting a bottom-up public-driven activity in the communities. From our perspective, the most promising results of CAP operation were obtained in developing of personal systems of values. This task was set in Equity Theory (Adams, 1963), but it was difficult to implement due to the imperceptibility of the "value" concept.

Modern semantically-enabled CAPs functionality allows users to transparently record not only own academic achievements, but also personal systems of values (PSV) in terms of flexible multidimensional and multi-contextual quality indicators weighed by their importance. Academic achievements of organizations (or their members) evaluated according to the PSVs produce personal ranking lists. But, what is more important, this mechanism generates better understanding of the complex web of relationships in academic communities by elicitation of: personal value of results and rewards; expert opinions and best practices (for benchmarking); change agents; differences in values systems (for conflict management) etc.

The first composed system of values during the TRUST piloting in NURE revealed the lack of academic culture within university community and showed a considerable conflict in understanding of academic values. The TRUST Portal played the educational role for the essential part of NURE academic staff introducing internationally recognized impact-valued systems. Benchmarking

---

[2] https://www.jyu.fi/coin/

of leaders (change agents) caused considerable transformations in portfolios of NURE academics.

Implementation of the participative decision-making concept is also of great interest (Cotton et al., 1988). Variety of methods for group decision-making (Hwang & Lin, 2012) create endless possibilities for collective value systems (CSVs) modelling.

As our experiments showed, one should be careful with the "wisdom of crowd", when CSV is defined as a simple average of PSVs. The crowd is far from being infallible, moreover, there is an undermining effect of social influence, which creates a misplaced bias (Lorenz et al., 2011). In our case the average CSV rejected the most impressive results. That`s why the HR-model "League system" was developed in NURE for self-development motivation (Semenets et al., 2021). It is based on ranking achievements according to the beforehand published PSVs and it works similarly to football regulations: three segments - leagues with different options in the administrative ranking list: the premier league (Seniors, those who make decisions), the first one (Middles, objects supported for capacity building) and the second one (Juniors, reserve for "ballast discharge"); the top-ranked person as the group`s leader; recalculation of ranks according to the new leaders` PSVs; reconfiguration for social elevator promotion (OECD, 2018): three top-ranked persons from a lower league replace three persons with the lowest rank.

This model is still being tested as an element of NURE reform. The case of TRUST Portal shows how collective intelligence and IR tools can bring change to the conservative, resisting and antagonizing university community and help starting a process of evolution towards internationally recognized quality and trust.